\documentclass{mem}
\usepackage{natbib}\usepackage{txfonts}\usepackage{balance}
\usepackage{graphicx}
\usepackage[a4paper]{hyperref}
\idline{79}{1}
\begin{document}

\title{
Cosmological evolution of supermassive black holes and AGN: a
synthesis model for accretion and feedback
}

   \subtitle{}

\author{
A. \,Merloni\inst{1,2}
          }

  \offprints{A. Merloni}

\institute{
Excellence Cluster Universe,
Boltzmannstr. 2,
85748, Garching, Germany
\and
Max Planck Institute f{\"u}r Extraterrestrische Physik,
Giessenbachstr.,
85741, Garching, Germany
\email{am@mpe.mpg.de}
}

\authorrunning{Merloni }

\titlerunning{Cosmological Evolution of SMBH}

\abstract{
The growth of supermassive black holes (SMBH) through accretion is
accompanied by the release of enormous amounts of energy which can
either be radiated away, as happens in quasars, advected into the
black hole, or disposed of in kinetic form through powerful jets, as
is observed, for example, in radio galaxies. Here, I will present new constraints
on the evolution of the SMBH mass function and Eddington ratio distribution,
obtained from a study of AGN luminosity functions aimed at accounting
for both radiative and kinetic energy output of AGN in a systematic way.
First, I discuss how a refined Soltan argument leads to
joint constraints on the mass-weighted average spin of SMBH and of the
total mass density of high redshift ($z\sim 5$) and ``wandering''
black holes. Then, I will show how to describe the 
``downsizing'' trend observed in the  AGN
population in terms of cosmological evolution of physical quantities 
(black hole mass, accretion rate, radiative and kinetic energy output).
Finally, the redshift evolution of the AGN kinetic
feedback will be briefly discussed and compared with the radiative
output of the evolving SMBH population, thus providing a robust
physical framework for 
phenomenological models of AGN feedback within structure formation. 
\keywords{accretion, accretion discs - black hole physics - galaxies:
  active - galaxies: evolution - quasars: general}
}
\maketitle{}

\section{Introduction: a synthetic picture of AGN evolution}

\begin{figure*}[t!]
\resizebox{\hsize}{!}{\includegraphics[clip=true]{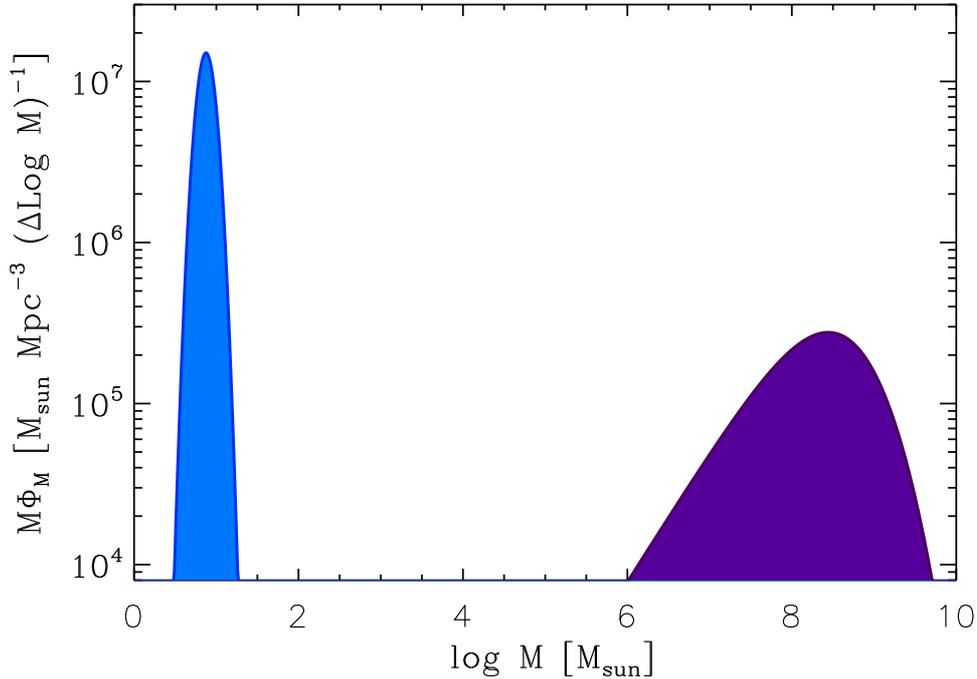}}
\caption{
\footnotesize
The local black hole mass function, plotted as $M \times \phi_M$, in
order to highlight the location and height of the two main peaks. The
stellar mass black holes peak has been drawn assuming a log-normal
distribution with mean mass equal to 5 solar masses, width of 0.1 dex
and a normalization yielding a density of about $1.1 \times 10^{7}$
$M_{\odot}$ Mpc$^{-3}$ \citep{fukugita:04}. 
The supermassive black hole peak, instead,
contribute to an overall density of about $4.3 \times 10^{5}$
$M_{\odot}$ Mpc$^{-3}$ \citep{merloni:08}}
\label{fig:mf}
\end{figure*}

Black holes in the local universe come into two main families
according to their size, as
recognized by the strongly bi-modal distribution of the local black
hole mass function (see Fig.~\ref{fig:mf}). While the height, width
and exact mass scale of the stellar mass
peak should be understood as a by-product of stellar (and binary)
evolution, and of the physical processes that make supernovae
and gamma-ray bursts explode, the supermassive black holes one is the
outcome of the cosmological growth of structures and of the evolution
of accretion in the nuclei of galaxies,
likely modulated by the mergers these nuclear black holes will
experience as a result of the hierarchical galaxy-galaxy coalescences.

In the recent  literature, it has become customary  to introduce 
the works on cosmological aspects of AGN astrophysics by referring to the strong
role they most likely play in the galaxy formation process throughout
cosmic history. Indeed, a new paradigm has emerged, according to which the
feedback energy released by growing supermassive black holes
(i.e. AGN) limits the stellar mass growth of their host galaxies in a
fundamental, generic, but yet not fully understood fashion.

The strongest observational evidence for such a schematic picture 
emerged in the last decade. 
The search for the local QSO relics via the study of their dynamical
influence on the surrounding stars and gas carried out since the launch of
the {\it Hubble Space Telescope} \citep[see
e.g.][and references therein]{richstone:98,ferrarese:08} 
led ultimately to the discovery of tight scaling relations
between SMBH masses and properties of the host galaxies' bulges
\citep{gebhardt:00,ferrarese:00,tremaine:02,marconi:03},
clearly pointing to an early co-eval stage of SMBH and galaxy growth.
A second piece of evidence comes from X-ray observations of galaxy clusters,
showing that black holes are able to deposit large amounts of energy into their
environment in 
response to radiative losses of the cluster gas. From studies of 
the cavities, bubbles and weak shocks
  generated by the radio emitting jets in the intra-cluster medium
  (ICM) it appears that AGN are
 energetically able to balance radiative losses from the ICM
 in the majority of cases \citep[see][and references
 therein]{birzan:08}.

Nevertheless, the physics of AGN heating in galaxy cluster is still not
well established, neither have the local scaling relations 
proved themselves capable to
uniquely determine the physical nature of the SMBH-galaxy coupling.
As a consequence, a large number of feedback models have so far 
been proposed which can
reasonably well reproduce these relations. From the
observational point of view, the crucial test for most models will be
a direct comparison with the high-redshift evolution of the scaling
relations. 

There is, however,  another benchmark, based on existing
data, all models have to be tested upon: the evolution of the SMBH
mass function and of the predicted energy output (either in radiative
or kinetic form) needed to offset gas cooling and star formation in galaxies.

Here I present our recent attempt to
reconstruct the history of SMBH accretion in order to follow closely 
the evolution of the black hole mass function, as needed in order to test
various models for SMBH cosmological growth as well as those for the black
hole-galaxy co-evolution. 
Similar to the case of X-ray background synthesis models, 
where accurate determinations of the XRB intensity and spectral shape, 
coupled with the resolution of this radiation into
individual sources, allow very sensitive tests of how the AGN luminosity
and obscuration evolve with redshift, 
we have argued that accurate determinations
of the local SMBH mass density and of the AGN (bolometric) luminosity functions,
coupled with accretion models that
specify how the observed AGN radiation translates into a black hole
growth rate, allow sensitive tests of how the SMBH population (its mass
function) evolves with redshift. By analogy, we have named
this exercises `AGN synthesis modelling' \citep{merloni:08}. 
In performing it, we have
taken advantage of the fact that the cosmological evolution of SMBH is
markedly simpler than that of their host galaxies, as individual black
hole masses can only grow with time, and SMBH do not transform into
something else as they grow. Moreover, by identifying active AGN
phases with phases of black holes growth, we can follow the 
evolution of the population by solving a simple
continuity equation, where the mass
function of SMBH at any given time can be used to predict that at any
other time, provided the distribution of accretion rates as a function
of black hole mass is known (see \S~\ref{sec:integral}).

\begin{figure*}[t!]
\begin{tabular}{cc}
\resizebox{0.48\hsize}{!}{\includegraphics[clip=true]{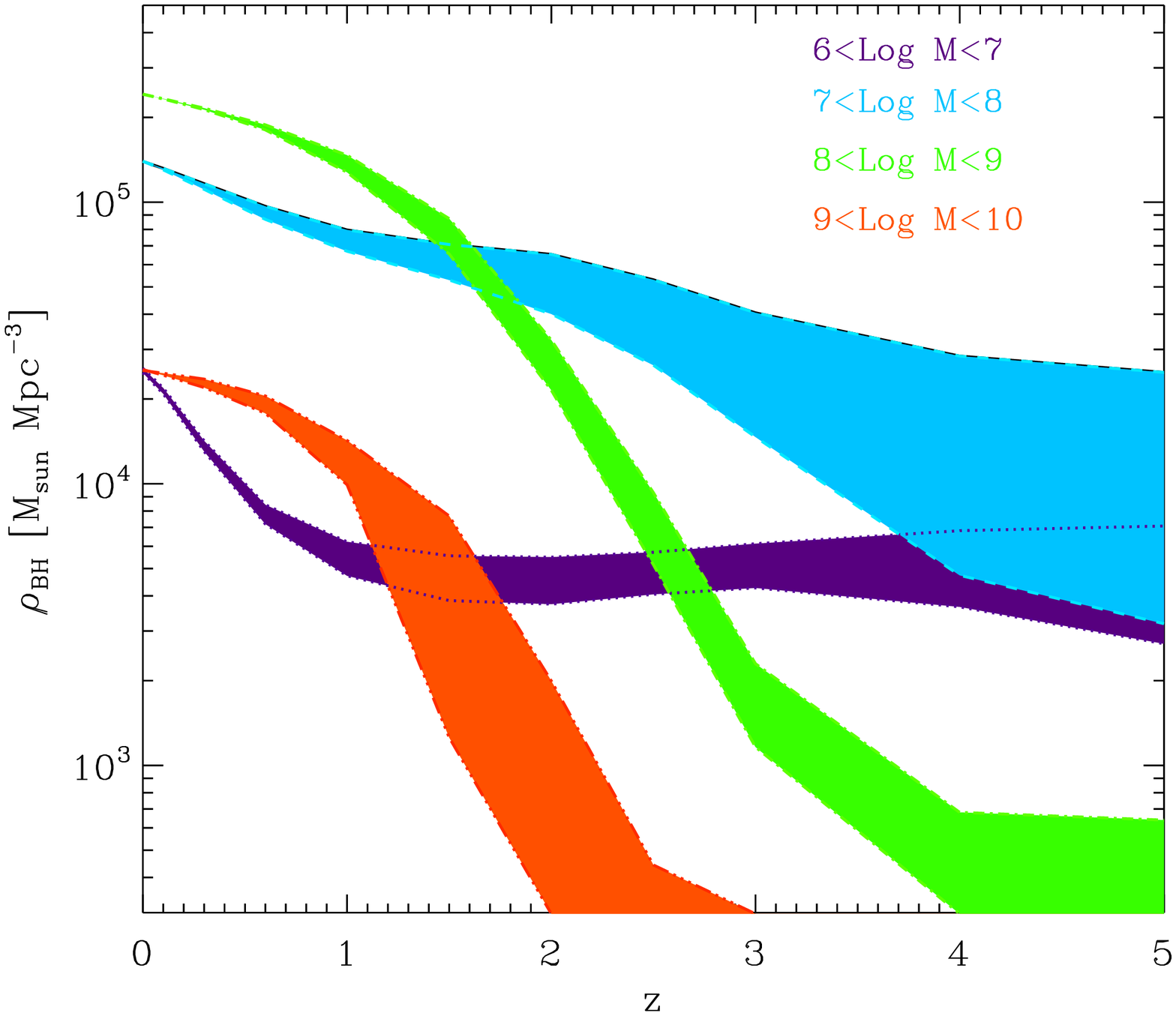}}&
\resizebox{0.48\hsize}{!}{\includegraphics[clip=true]{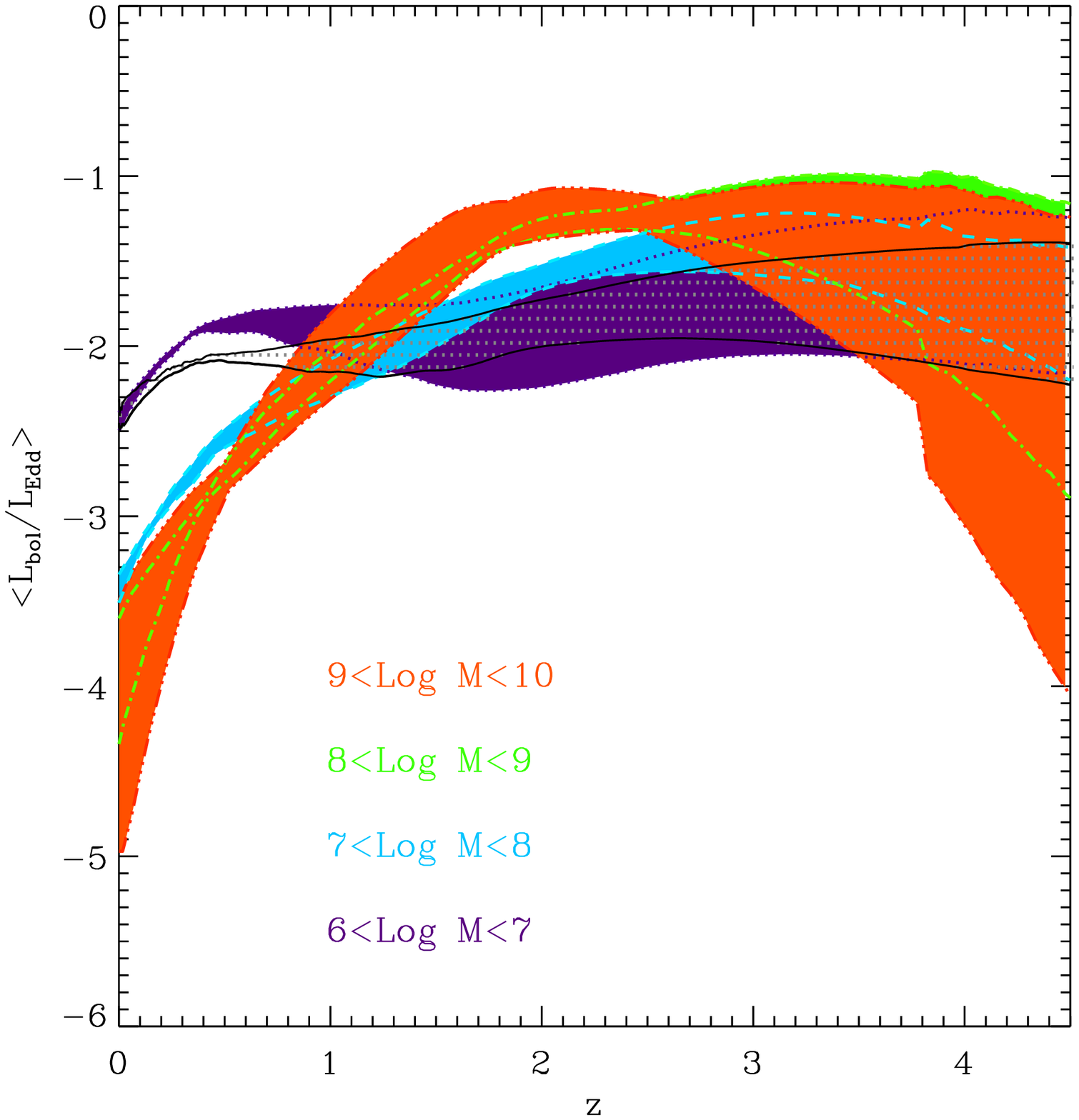}}\\
\end{tabular}
\caption{\footnotesize
Redshift evolution of the SMBH mass density (left) and average
Eddington ratio (right), calculated for different BH mass bins (in
solar mass units). In the
right hand plot, black lines and grey shaded area represent the
overall (mass-wighted) average Eddington ratio.}
\label{fig:zev_mdot}
\end{figure*}

In order to carry out our calculation, we assumed that black
holes accrete in just three distinct physical states, or ``modes'':
a radiatively inefficient, 
kinetically dominated mode at low Eddington ratios (LK, the so-called
``radio mode'' of the recent literature), and two modes at 
high Eddington ratios: a purely radiative one (radio quiet,
HR), and a kinetic (radio loud, HK) mode, with the former outnumbering
the latter by about a factor of 10. Such a classification is based on
our current knowledge of state transitions in stellar mass black hole
X-ray binaries as well as on a substantial body of works on scaling
relations in nearby SMBH. It allows a
relatively simple mapping of the observed luminosities (radio cores,
X-ray and/or bolometric) into the physical quantities related to any
growing black hole: its accretion rate and the released kinetic power.

In this work I will focus on just a few specific
 aspects of the derived SMBH evolution, in particular on the redshift
 evolution of the mass function and Eddington ratio distribution, on
 the constrains we put on the mass-weighted average spin of the SMBH
 population, and on the kinetic energy output of growing black holes.
A more detailed discussion of the methodology, as well as a wider
exploration of our results can be found in \citet{merloni:08}.

All results will be shown
accounting for the intrinsic uncertainties of the adopted luminosity
functions. We estimated that these uncertainties can be evaluated by
comparing different analytic parametrization of the same data sets;
specifically, we adopted 
the LDDE and MPLE parametrization for the hard X-ray luminosity function of
\citet{silverman:08}, and two alternative
parametrizations for the flat-spectrum radio luminosity function of
\citet{dunlop:90} and \citet{dezotti:05}.

\section{The evolution of SMBH Eddington ratio}
\label{sec:integral}

We studied the evolution of SMBH mass function through a continuity
equation that can be written as:
\begin{equation}
\label{eq:continuity}
\frac{\partial \psi(M,t)}{\partial t} +
\frac{\partial}{\partial M}\left( \psi(M,t) \langle \dot M
  (M,t)\rangle \right)=0
\end{equation}
where $M$ is the black hole mass, $\mu=Log\, M$, $\dot \mu=Log\, \dot M$, 
$\psi(M,t)$ is the SMBH mass function at time $t$, and $\langle \dot M
(M,t) \rangle$ is the average accretion rate of SMBH of mass $M$ at
time $t$, and can be defined through a ``fueling'' function,
$F(\dot\mu,\mu,t)$, describing the
distribution of accretion rates for objects of mass $M$ at time $t$: 
$\langle \dot M(M,z)\rangle = \int \dot M F(\dot\mu,\mu,z)\, \mathrm{d}\dot\mu$.

Such a fueling function is not a priori known, 
but can be derived inverting the integral
equation that relates the luminosity function of the population in
question with its mass function:
\begin{equation}
\label{eq:filter}
\phi(\ell,t)=\int F(\dot\mu,\mu,t) \psi(\mu,t)\; \mathrm{d}\mu
\end{equation}
where I have called $\ell=Log\, L_{\rm bol}$.

Thus, we have integrated eq~(\ref{eq:continuity}) starting from $z=0$, where
we have simultaneous knowledge of both mass, $\psi(M)$, and
luminosity, $\phi(\ell)$, functions,
evolving the SMBH mass function backwards in time, up to where
reliable estimates of the (hard X-ray selected) 
AGN luminosity functions are available.
The adopted hard X-ray luminosity function \citep{silverman:08} is
supplemented with luminosity-dependent bolometric
corrections \citep{marconi:04} 
and absorbing column density distributions 
consistent with the X-ray background  constraints (as
well as the sources number counts, and many others), following
the most recent XRB synthesis model \citep[see][for
details]{gilli:07}. 

In Figure~\ref{fig:zev_mdot}, I show the evolution of the black hole
mass density (left hand side) and of the mass-weighted average
Eddington ratio, $\lambda\equiv L_{\rm bol}/L_{\rm Edd}$ (right hand
side), both computed for four different black hole mass bins (in solar
mass units).

Between redshift zero and one, it is evident how small mass object
have a higher Eddington ratio, and increase their total density much
more rapidly than their high-mass counterpart, an effect of the well
known phenomenon called ``AGN downsizing'' \citep{heckman:04}. This
trend seems however inverted at higher redshift, when the largest
black holes are assembled; a more precise assessment of this
phenomenon will require better statistics on the high redshift AGN
luminosity functions.

\section{Constraints on the average black hole spin}

\begin{figure*}[t!]
\begin{tabular}{cc}
\resizebox{0.48\hsize}{!}{\includegraphics[clip=true]{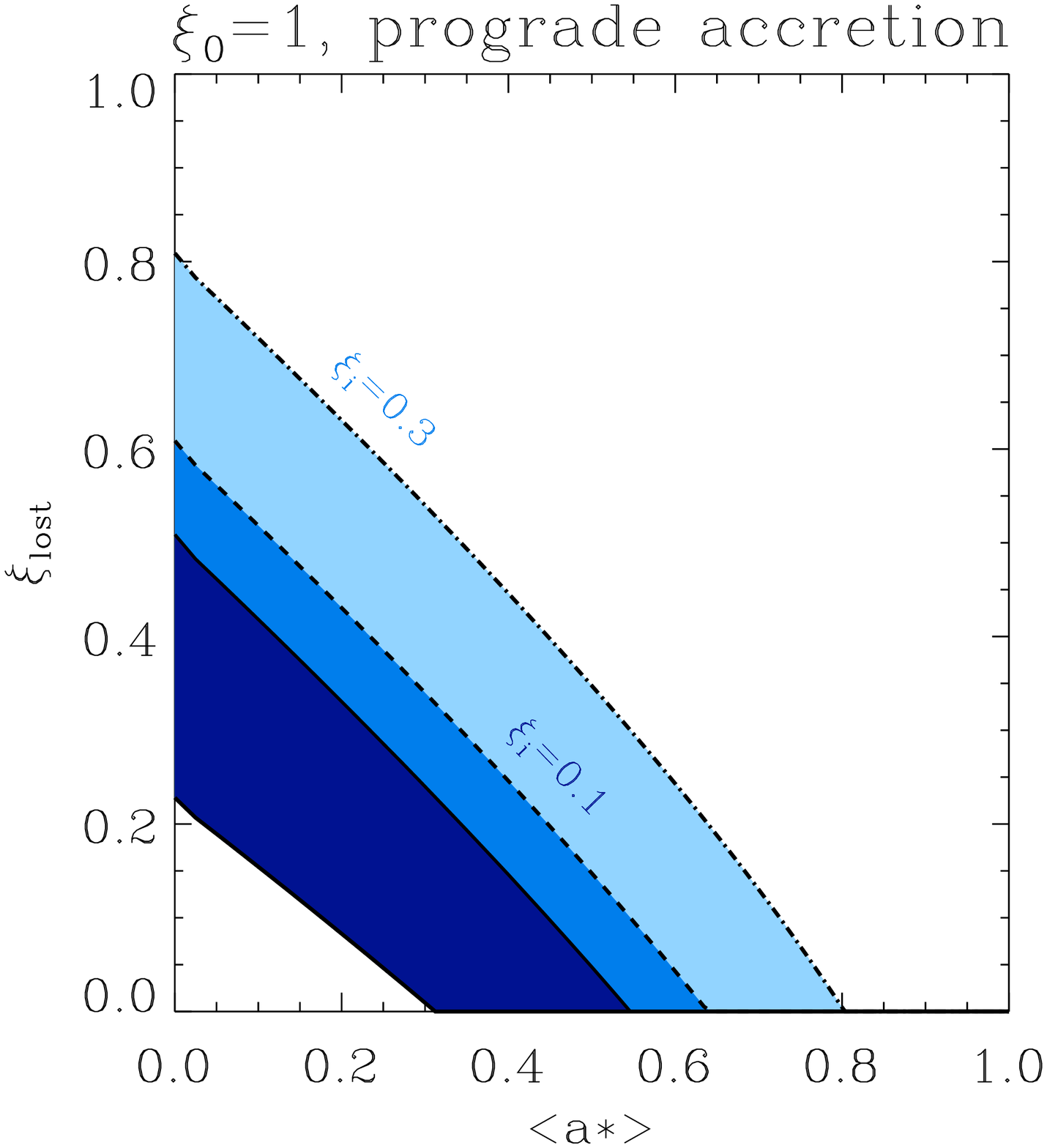}}&
\resizebox{0.48\hsize}{!}{\includegraphics[clip=true]{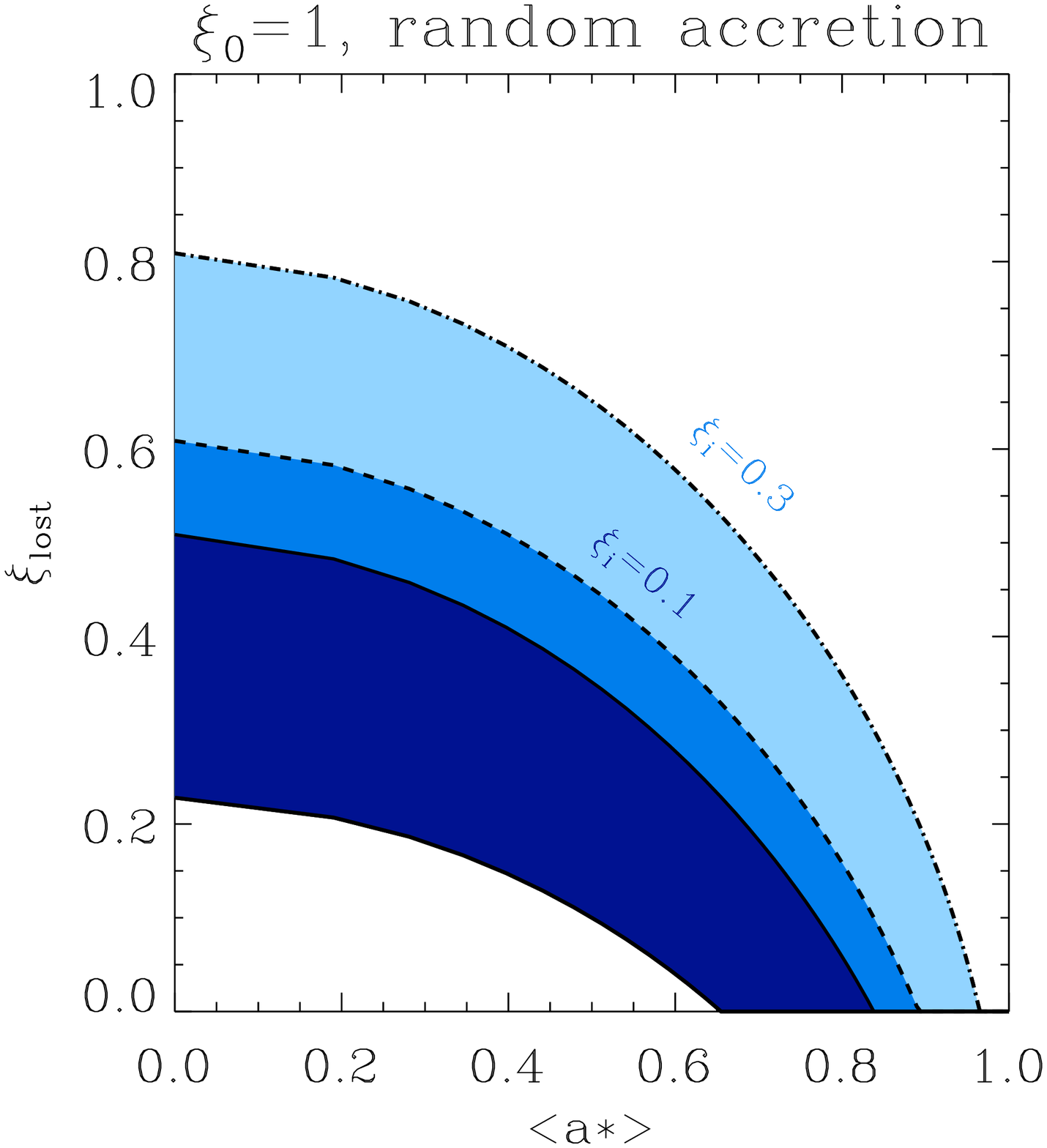}}\\
\end{tabular}
\caption{\footnotesize
Constraints on the mass-weighted average spin of SMBH and on the
fraction of ''wandering'' black hole mass density $\xi_{\rm
  lost}$. The dark blue contours between solid lines correspond to the
case of negligible black hole mass density at $z=5$ ($\xi_i=0$),
progressively lighter blue areas represent the case of larger and larger
values of $\xi_i$. The left hand side shows the calculation assuming
accretion onto SMBH proceed always in prograde fashion; the right hand
side shows the opposite extreme of purely random accretion. These two
extreme cases encompass all possible cosmological solutions.}
\label{fig:xi_a}
\end{figure*}

One of the most far-reaching conclusions of the analysis presented in
Merloni and Heinz (2008) was the necessity of a broad distribution of
Eddington ratio for SMBH of various masses and redshift,
in order to reconcile the observed evolution of the AGN luminosity
functions with that of the mass function itself. Consequently, there
are always black holes of any mass accreting in different ``modes'' at
any time, and this bears some consequences on the final estimates of
the average efficiencies.

In fact, in our three-mode scheme for black hole accretion, the radiative
efficiency, $\epsilon_{\rm rad}$, is not necessarily equal to the accretion
efficiency, $\eta$.  The
latter represents the maximal amount of
potential energy that can be extracted, per unit rest mass energy,
from matter accreting onto the black hole. This quantity, $\eta(a)$,
depends on the inner boundary condition of the accretion flow only,
and, in the classical no-torque inner boundary, is a
function of BH spin, $a$, only, ranging from $\eta(a=0)\simeq 0.057$ for
Schwarzschild (non-spinning) black holes to $\eta(a=1)\simeq 0.42$ for
maximally rotating Kerr black holes. 

On the other hand, the radiative efficiency, $\epsilon_{\rm rad}\equiv
L_{\rm bol}/\dot M c^2$ depends both on the accretion efficiency
(i.e. on the inner boundary condition of the accretion flow) and on
the nature of the accretion flow itself. Based on our current
knowledge of the physical properties of low and high luminosity AGN
(and stellar mass black holes), in \citet{merloni:08} we have adopted
the following parametrization:
\begin{equation}
\label{eq:radeff}
\epsilon_{\rm rad} \equiv \eta f(\dot m) =\eta \times \left\{
        \begin{array}{ll}
        1,   &  \dot m \ge \dot m_{\rm cr}   \\
        (\dot m/\dot m_{\rm cr}), & 
\dot m < \dot m_{\rm cr}   \\
        \end{array}\right.\;
\end{equation}
where $\dot m_{\rm cr}=\lambda_{\rm cr}$ is the critical 
Eddington-scaled accretion rate above which the disc becomes 
radiatively efficient, assumed, in our computation, to be at the
universal value of 0.03.

In \citet{merloni:08} we have shown how using
standard \citet{soltan:82} type of arguments, i.e. comparing the local
mass density to the integrated mass growth in AGN phases, very tight
constraints can be put on the average radiative efficiency of the
accretion process: 

\begin{equation}
\frac{0.065}{\xi_0(1+\xi_{\rm lost})}\la
\epsilon_{\rm rad} 
 \la \frac{0.070}{\xi_0(1-\xi_i+\xi_{\rm lost})},
\end{equation} 
where $\xi_0$ is the local
mass density in units of 4.3 $\times 10^{5} M_{\odot}$ Mpc$^{-3}$,
while $\xi_i$ and $\xi_{\rm lost}$ are the mass density of $z\sim 5$ and
``wandering'' SMBH, respectively (also in units of the local mass
density).

In order to translate our constrains on the radiative efficiency onto a
constrain on the average black hole spin, however, we need to
carefully consider the distribution of angular momentum of the matter
accreting onto the black holes. As discussed in \citet{king:08}, 
the most important factor is the amount of mass that
can be accreted ``coherently'', i.e. keeping the same large scale
angular momentum. If this is larger than a few per cent of the black
hole's mass, the black hole will be always spun up, and 
 most of the SMBH mass is accreted from a prograde disk. If, on
the other hand, accretion proceeds via small, independent (randomly
oriented) sub-units, prograde and retrograde accretion events are almost
equally probable \citep{king:08}, and we should use a ``symmetrized''
relation between black hole spin and accretion efficiency:
$\eta_{\pm}(a)=(1/2)[\eta(a)+\eta(-a)]$.
Most likely, the true average relation between spin and efficiency
will lie somewhere in between these two extreme cases, depending on
the detail of the SMBH fueling mechanism at various redshifts.

Figure~\ref{fig:xi_a} shows the constraints derived 
in the $\xi_{\rm lost}$, $\langle
a*\rangle$ plane, where $\langle a*\rangle$ is the mass weighted
average spin parameter of the SMBH calculated inverting either the classical
GR $\eta(a)$ relation (\citet{shapiro:05}; left panel, prograde accretion
only) and its symmetrized version (right panel, random accretion),
and taking into account the
assumed relationship of eq.~(\ref{eq:radeff}) between $\eta$ and
$\epsilon_{\rm rad}$.

It is interesting to note that both the amount (numbers and masses) of
black holes effectively ejected from galactic nuclei due to
gravitational wave recoil after a merger and the average radiative
efficiency of accreting black holes depend on the spin distribution of
evolving SMBH \citep[see e.g.][]{berti:08}. 
Thus, the plots of Fig.~\ref{fig:xi_a} couple implicitly the spin
distribution of accreting black holes (through $\xi_{\rm lost}$ and
$\epsilon_{\rm rad}$) and the properties of the seed
black hole population, whose density must be reflected in the $z\sim
5$ mass density $\xi_i$.

\section{The kinetic energy density output of SMBH}
\label{sec:kinetic}
Finally, I discuss here briefly the consequences of our synthetic
picture of AGN evolution for the issue of AGN feedback in the form of
kinetic energy associated to radio jets.

In order to do that, we start from the recently found correlation
between kinetic power and radio core luminosity of AGN jets 
\citep{merloni:07}, and use the flat spectrum radio luminosity function
evolution \citep{dunlop:90,dezotti:05} to derive the evolution of the
kinetic luminosity function of AGN. 
The effects of beaming have been taken into account statistically, both
in the core-kinetic  luminosity relation \citep{merloni:07}
and in the evolving flat spectrum radio luminosity functions \citep{merloni:08}.

\begin{figure*}[t!]
\begin{tabular}{cc}
\resizebox{0.48\hsize}{!}{\includegraphics[clip=true]{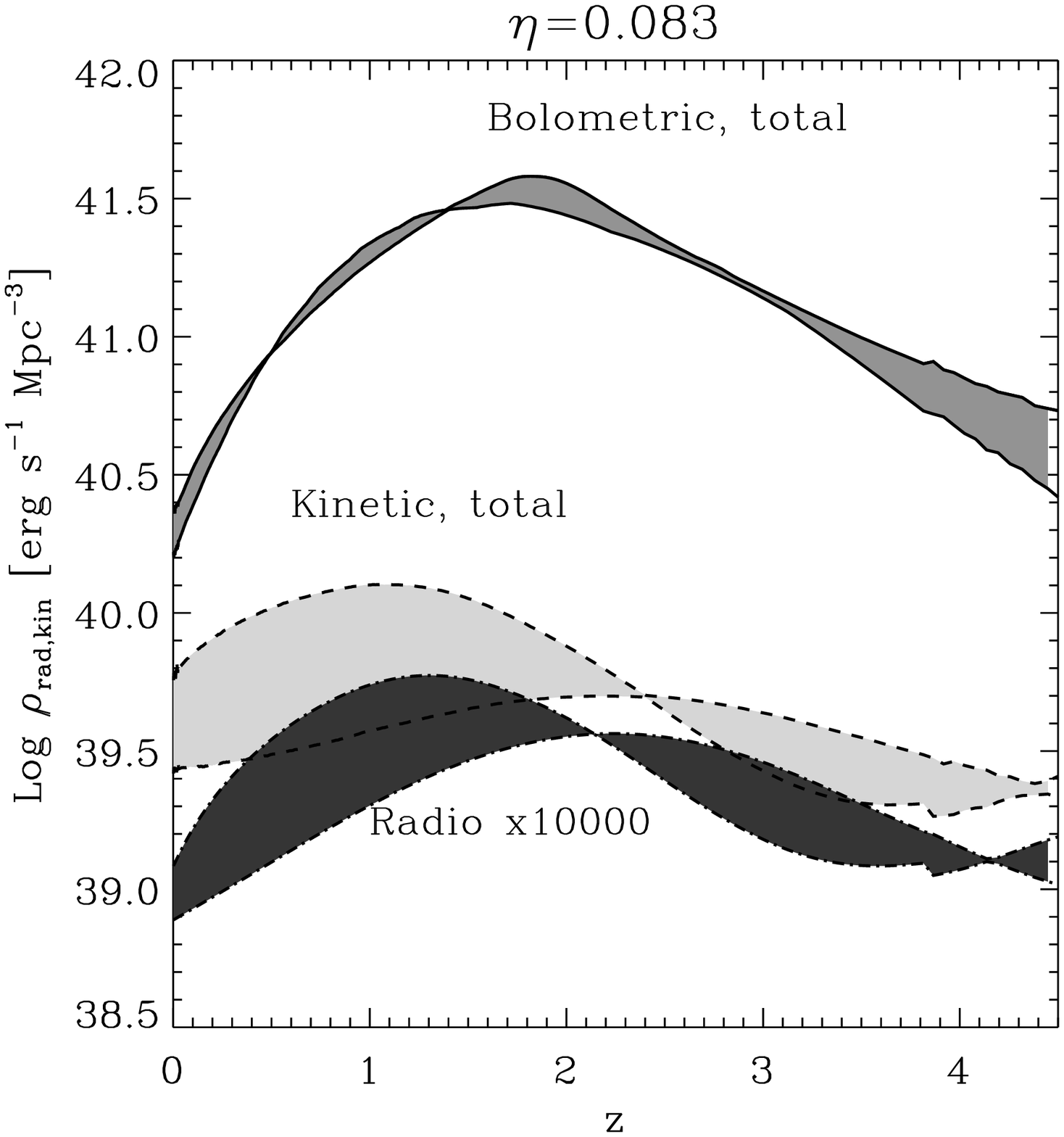}}&
\resizebox{0.48\hsize}{!}{\includegraphics[clip=true]{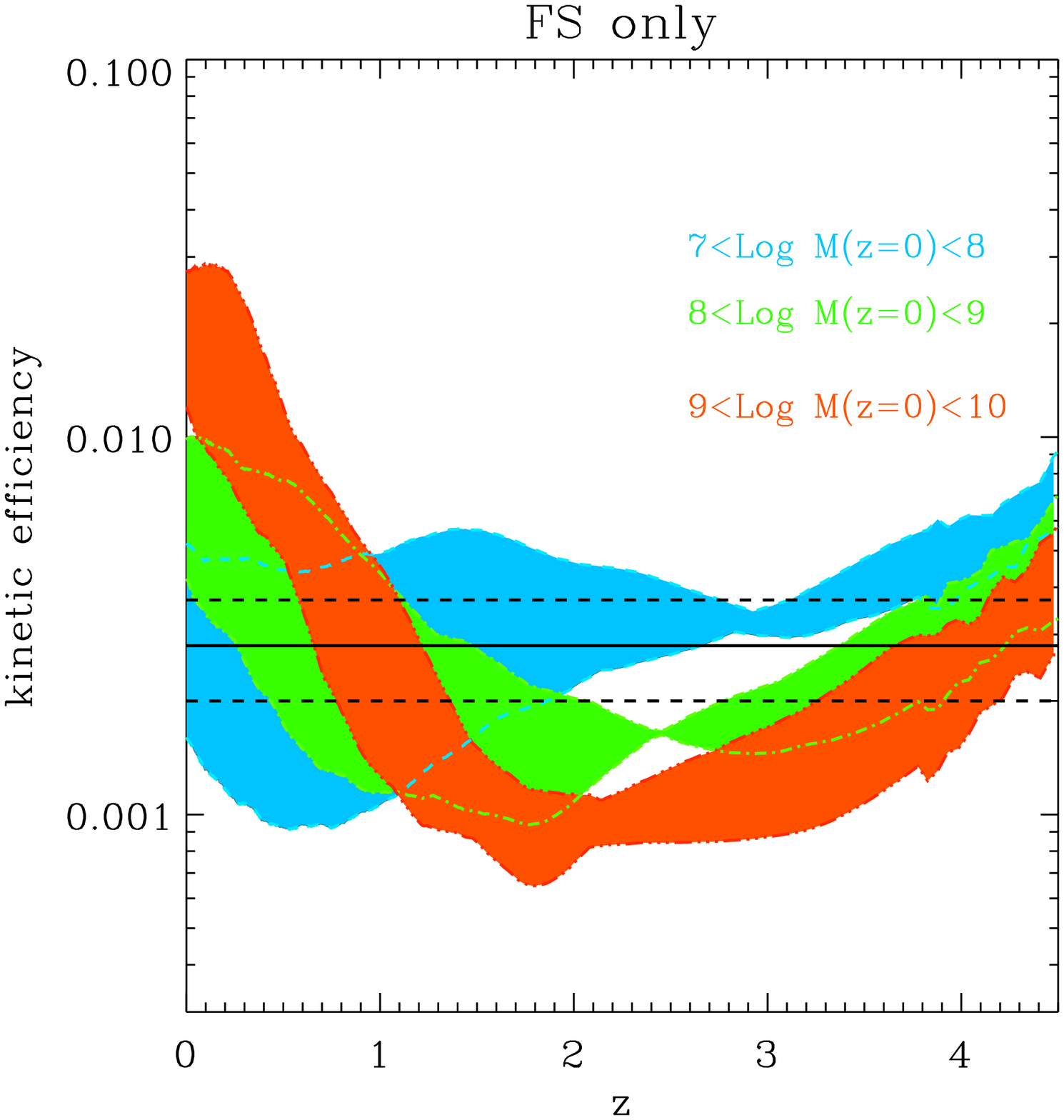}}\\
\end{tabular}
\caption{\footnotesize
Left: the AGN kinetic energy density output as a function of
redshift,
(light grey area between dashed lines) 
compared to the (bolometric) radiation energy
density output (grey area between solid lines) and the radiation
energy density in the radio band (5 GHz; dark grey area between
dot-dashed lines). Right: Redshift evolution of the kinetic
efficiency. 
SMBH of different masses at $z=0$ are here plotted separately, with
  a color coding analogous to that of Fig.~\ref{fig:zev_mdot}; 
the horizontal black solid lines mark the mass weighted average
values for the kinetic efficiencies, with the dashed lines
representing the uncertainties from the particular choice of radio and
X-ray luminosity functions (see Merloni and Heinz 2008 for more details).}
\label{fig:zev_kin}
\end{figure*}

By integrating over the kinetic luminosity functions, we get an
estimate of the AGN kinetic energy density output as a function of
redshift, as shown in the left hand panel of Fig.~\ref{fig:zev_kin},
where a comparison is made with the (bolometric) radiation energy
density output and the radiation
energy density in the radio band (5 GHz).  
The most notable feature of this plot is the markedly different
redshift evolution of the radiative and kinetic power output, with the
latter showing a much smaller amount of evolution between $z=0$ and
$z=1$. This is due essentially to the increasing number of SMBH
accreting at low Eddington ratios in the so-called ``low-kinetic''
(LK) mode, where a substantial fraction of the gravitational energy of
the accreted mass is converted into jet kinetic power. Such a weak redshift
dependence, already suggested by \citet{merloni:04}, is in fact necessary
in order to reproduce the high-end of the galaxy mass function in
semi-analytic models of structure formation that invoke AGN feedback
\citep{croton:06,bower:06}.

Let us now examine the
 kinetic energy production efficiency of
 growing black holes.
We first compute the total integrated (mass
 weighted) average kinetic efficiency as\footnote{For the sake of
   simplicity, I discuss here only the case in which we derive the
   intrinsic radio core luminosity function from the flat spectrum
   sources only; an upper limit to the contribution from hidden cores
   in steep spectrum radio sources have been considered in Merloni and
 Heinz (2008), yielding a final upper limit for the kinetic efficiency
about a factor of 2 higher}:
\begin{equation}
\label{eq:eps_kin}
\langle\epsilon_{\rm kin}\rangle \equiv \frac {\int_0^{z_i} \rho_{\rm
    kin}(z) dz}{\int_0^{z_i}c^2 \Psi_{\dot M}(z) dz} \simeq (2.8\pm
0.8) \times 10^{-3}
\end{equation}
where $\Psi_{\dot M}(z)$ is the total black hole accretion rate
density at redshift 
$z$. Combining the results discussed in section~\ref{sec:integral} on the
average radiative efficiency with those shown in
eq.~(\ref{eq:eps_kin}) we conclude that SMBH, during their growth from
$z\sim 5$ till now, convert about 25 (down to 15 if hidden cores are
considered) times more rest mass energy into
radiation than into kinetic power, with the exact number depending on
the poorly known details of the intrinsic jet cores luminosity
function, as well as on our assumptions about the beaming corrections
to be made. Similar, but complementary studies of the AGN kinetic
luminosity function evolution based instead 
on {\it steep} spectrum sources has been
recently carried out \cite{koerding:08,shankar:08}. Reassuringly, 
the results of both
these works are in reasonable agreement with those I presented here.
 
We can also compute directly the kinetic efficiency as a
function of redshift and SMBH mass {\it today}, which I show in the right hand
panel of Fig.~\ref{fig:zev_kin}, with the horizontal
solid lines showing the mass weighted average from
eq.~(\ref{eq:eps_kin}). 
The various curves describe the main
properties of kinetic feedback as we observe it. 
For each of the chosen mass ranges,
the kinetic efficiency has a minimum when black holes of that mass
experience their fastest growth: 
this is a different way of restating the conclusion that most
of the growth of a SMBH happens during radiatively efficient phases of
accretion. However, when the mass increases, SMBH are more and more
likely to enter the LK mode, which
increases their kinetic efficiency. More massive holes today, have entered this
phase earlier, and by $z=0$ they have reached the highest kinetic
efficiency of a few $\times 10^{-2}$. This is a natural consequence of
the observed anti-hierarchical growth of the SMBH population, and of
the chosen physical model for the accretion mode of low-Eddington
ratio objects.

\section{Conclusions}
I have outlined some recent results of our work aimed at 
pinning down as accurately as possible the cosmological evolution 
of active galactic nuclei and of the associated growth of the supermassive
black holes population. 

In particular, I have focused here on the global (integrated)
constraints on the mass-weighted average spin of SMBH, and I have
discussed in some details the specific ways in which these are tighten
to the very interesting open issues regarding the population of
high-redshift SMBH and that of black holes ejected from galactic
nuclei due to gravitational wave recoil in merger events.

I have also discussed a few generic properties of the kinetic energy
output of growing black holes, emphasizing the importance of a late
phase of low-luminosity, jet-dominated accretion onto the most massive
objects. 

The richness of details we have been able to unveil demonstrates that 
times are ripe for comprehensive unified approaches to
the multi-wavelength AGN phenomenology. At the same time, our results
should serve as a stimulus for semi-analytic and numerical modelers of
structure formation in the Universe to consider more detailed
physical models for the evolution of the black hole population.

\begin{acknowledgements}
I would like to thank warmly Sebastian Heinz for the fruitful
collaboration over the last few years, and the organizers of the meeting "The
central Kiloparsec: Active Galactic Nuclei and their Hosts" for the
hospitality and the stimulating meeting they have made possible.
\end{acknowledgements}

\bibliographystyle{aa}

\end{document}